\documentclass{emulateapj}

\shorttitle{Multi-ion plasma turbulence in the solar wind}
\shortauthors{Perrone et al.}

\begin{document}

\title{Vlasov simulations of multi-ion plasma turbulence in the solar wind}

\author{D. Perrone$^{1}$, F. Valentini$^{1}$, S. Servidio$^{1}$, S. Dalena$^{1,2}$, and P. Veltri$^{1}$}
\affil{
${1}$ Dipartimento di Fisica and CNISM, Universit\`a della Calabria, I-87030 Rende (CS), Italy \\
${2}$ Bartol Research Institute, Department of Physics and Astronomy, University of Delaware, Newark, DE 19716, USA}

\begin{abstract}
Hybrid Vlasov-Maxwell simulations are employed to investigate the role of kinetic effects
in a two-dimensional turbulent multi-ion plasma, composed of protons, alpha particles and
fluid electrons. In the typical conditions of the solar-wind environment, and in 
situations of decaying turbulence, the numerical results show that the velocity
distribution functions of both ion species depart from the typical configuration of thermal equilibrium.
These non-Maxwellian features are quantified through the statistical analysis of the temperature anisotropy, for both protons and 
alpha particles, in the reference frame given by the local magnetic field. Anisotropy 
is found to be higher in regions of high magnetic stress. Both ion species manifest a preferentially perpendicular heating, although the anisotropy is more pronounced for the alpha particles, according with solar wind observations. 
Anisotropy of the alpha particle, moreover, is correlated to the proton anisotropy, and also depends on the local 
differential flow between the two species. Evident distortions of the particle distribution functions are present, with the production of bumps along the direction of the local magnetic field. The physical phenomenology recovered in these numerical simulations reproduces very common measurements in the turbulent solar wind, suggesting that the multi-ion Vlasov model constitutes a valid approach to the understanding of the nature of complex kinetic effects in astrophysical plasmas.
\end{abstract}

\section{Introduction}
Astrophysical plasmas are generally in a fully turbulent regime. In particular,
the solar wind can be considered as a natural laboratory for studying physical
processes of plasma turbulence, whose dynamic scales cannot be achieved in
laboratory experiments. Collisions in the solar wind are very rare: the particle mean free path is
comparable to (or larger than) the system size ($\simeq 1$ AU). These properties
suggest that small-scale mechanisms should be more complex than one would expect
from a fluid (viscous) description.

Spacecraft measurements generally reveal that the electromagnetic field fluctuations in the solar wind are in a state
of fully-developed turbulence \cite[]{bru05, mar06}. The power spectrum of the fluctuating fields manifests
a behavior reminiscent of the $k^{-5/3}$ power-law for fluid turbulence \cite[]{kol41}, where $k$ here is the wavenumber 
obtained applying the Taylor hypothesis. This inertial range turbulence extends to smaller spatial scales down to a range of wavelengths where kinetic effects dominate the plasma dynamics. At these scales, different physical processes come into play, leading to spectral shape changes. The first clear spectral change appears at scales such as the ion inertial length and/or the ion Larmor radius \cite[]{leamon00,bourouaine12,bal05}, where the spectrum of the magnetic field gets steeper \cite[]{bal05}. At the range of spatial lengths of the order of the electron kinetic scales the interpretation of the solar-wind observations is still controversial. Two different scenarios have been recently pictured: a second spectral break with an additional power-law range \cite[]{sah10} and an exponential cut-off \cite[]{ale09} that instead marks the end of self-similarity. In both cases, a theoretical support from self-consistent, fully nonlinear, Vlasov models is needed for the interpretation of this complex phenomenology.

This general picture of astrophysical turbulence becomes more complicated because of the multi-component nature of the solar wind. The interplanetary medium, although predominantly constituted of protons, is also made of a finite amount of doubled ionized helium (alpha particles), together with a few percentage of heavier ions.
Several observations \cite[]{mar82_a, mar82_b, kas08} have shown that heavier ions are heated and accelerated preferentially as compared to protons and electrons. Moreover, in recent analysis performed on solar-wind data from the Helios spacecraft, the link between the signatures of kinetic effects and some important parameters of heavy ions, such as relative speed, temperature ratio and anisotropy, has been investigated  \cite[]{bou10, bou11_a, bou11_b}. In these works the authors pointed out that more significant anisotropies and non-Maxwellian features are detected for alpha particles distribution functions with respect to protons. The evolution of the velocity distribution functions in the solar wind, and the production of kinetic signatures such as heating and temperature anisotropies represent nowadays some of the key issues of plasma physics \cite[]{osm10, osm12}.

The two points discussed above, that the kinetic scales have a determinant effect in shaping the turbulent spectra and that the role of secondary ions cannot be neglected, suggest that a multi-scale and multi-species self-consistent Vlasov treatment of the turbulent solar wind is required. The detection of the main kinetic processes that, acting on small scales, are responsible for the energy dissipation and the heating production, represents today one of the main challenges for the plasma scientific community. A kinetic description of collisionless plasma turbulence offers the powerful opportunity of giving important insights for the interpretation of ``in situ'' satellite measurements in the solar wind. In this context, an indispensable and crucial support to investigate the complexity of solar-wind
physics is represented by kinetic numerical simulations. 

The most widely adopted numerical description of the kinetic plasma physics is represented by the Particle in Cell (PIC) methods. The Lagrangian PIC approach is based on the integration of the equations of motion of a large number of macro-particles under the effects of self-consistent electromagnetic fields. The PIC simulations have been extensively used for the description of the kinetic dynamics of space plasmas, addressing the problem of wave-particle interaction \cite[]{ara08},
particle heating \cite[]{ara09} and turbulence \cite[]{gar08, sai08, par10, cam11, par11, mark2011}. Nowadays, thanks to the fast technological development of the computational 
resources, the Eulerian approach for the numerical solution of the Vlasov equation has become accessible as an alternative to the PIC approach \cite[]{man02}. The Eulerian Vlasov simulations, in which the time evolution of the particle distribution function is followed numerically in a discretized phase space domain, are significantly more demanding from the computational point of view compared to the PIC simulations. Nevertheless, at variance with the PIC methods, the Vlasov algorithms are mostly noise-free and do not introduce additional (unphysical) heating due to the particle noise. This point can be crucial when dealing with the numerical description of the short wavelengths part of the turbulent cascade, where the energy level of the fluctuations is typically very low and the statistical noise introduced by the PIC calculations can possibly cover the physical information.

Recently \cite[]{man02, val05, val07}, an Eulerian hybrid Vlasov-Maxwell code (HVM hereafter) has been developed. This algorithm integrates numerically the Vlasov equation in phase space coupled to the Maxwell equations for the electromagnetic fields. The Vlasov equation is solved for the proton species, while electrons are treated as a fluid \cite[]{val07}. The HMV code has been extensively used for the analysis of the kinetic effects during the evolution of the solar-wind cascade \cite[]{val08, val09, val10, val11}.
Furthermore, \cite{ser12} made use of the  HMV algorithm to investigate the role of local kinetic effects in plasma turbulence in a 2D-3V phase space configuration (two dimensions in physical space, and three in velocity space). It has been shown that, nearby the region of strong magnetic activity, the proton distribution function is deformed
by kinetic effects displaying significant non-Maxwellian features. Moreover, in these regions characterized by high magnetic stress,  reconnection events can occur locally, as it has been also described in both PIC and Magnetohydrodynamic simulations \cite[]{ser09,dra10}.

As discussed previously, in the solar wind description it is important to take account of its multi-component nature. For this purpose, \cite{per11} have proposed an update version of the HVM code, which includes the kinetic dynamics of heavy ions. In particular, the authors examined the effects produced by the presence of alpha particles
in the evolution of the solar-wind turbulent cascade in the direction parallel to the ambient magnetic field in a 1D-3V phase space configuration.

In the present paper we discuss the results obtained by the ``multi-component'' version of the HVM code in a 2D-3V phase space configuration. In order to study the turbulent activity in the presence of alpha particles, we adopt an approach similar to the one used by  \cite{ser12}, extending the above results to the more realistic multi-ion treatment. As we will discuss in the following, our numerical simulations reproduce several features commonly observed in space plasmas. As the result of the turbulent cascade, coherent structures appear in the system, with regions of high magnetic stress where reconnection locally occurs; in correspondence of these regions the particle distribution functions are heavily distorted, exhibiting a significant temperature anisotropy. This effect is more evident for alpha particles than for protons, consistent with recent observations in the solar wind \cite[]{bou10, bou11_b}.

\begin{figure}
\begin{center}
\includegraphics [width=8.4cm]{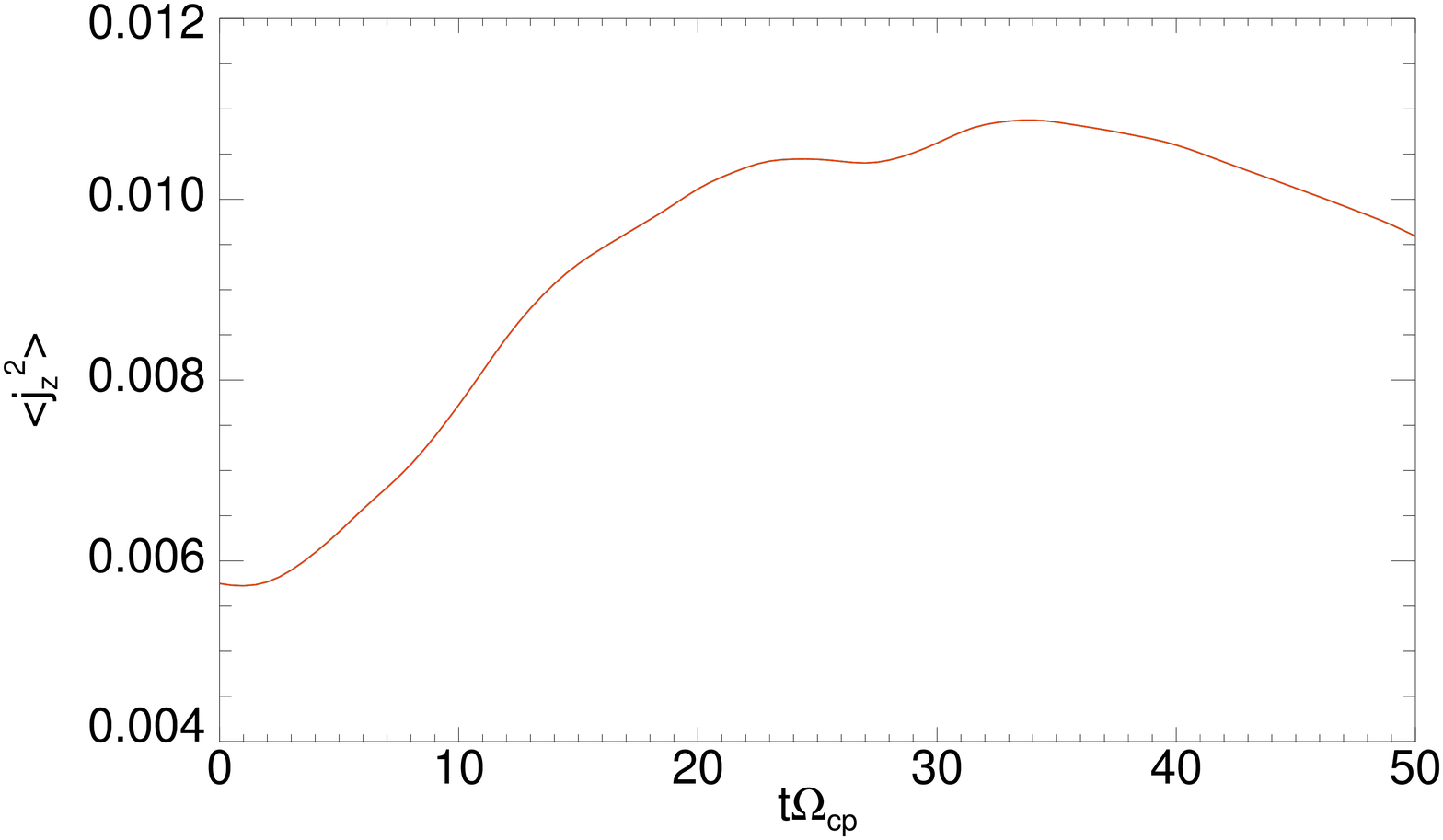}
\caption{Time evolution of the average out-of-plane squared current density $\langle j_z^2\rangle$.}
\label{fg:turbulence}
\end{center}
\end{figure}
\section{Numerical Model}

We simulate a collisionless and magnetized multi-species (electrons,
protons and alpha particles) turbulent plasma through the use of the HVM 
code. Within this HVM model, the Vlasov equation for proton
$(f_p)$ and alpha particle $(f_{\alpha})$ distribution functions \cite[]{val07,
per11} is integrated numerically in a 2D-3V phase space domain.
Electrons are treated as a fluid and a generalized Ohm equation, where a resistive term has been added as a standard numerical Laplacian dissipation, is considered. 
The dimensionless HVM equations are given by
\begin{eqnarray}
\label{eq:Vlasov}
& &\frac{\partial f_i}{\partial t} + {\bf v} \cdot \frac{\partial f_i}{\partial {\bf r}} + 
({\bf E} + {\bf v} \times {\bf B}) \cdot  \frac{\partial f_i}{\partial {\bf v}} = 0,\\
& &{\bf E} = -({\bf u}_e \times {\bf B})- \frac{1}{n_e} \nabla P_e + \eta {\bf j}\label{eq:ohm},\\
& & \frac{\partial {\bf B}}{\partial t}=-\nabla\times {\bf E}.
\label{eq:farampere}
\end{eqnarray}
The quasi-neutrality approximation $n_e=\sum_i Z_i n_i$ is considered, where $n_e$ and $n_i$ are the electron and ion densities respectively and $Z_i$ the ion charge number (the subscript $i=p,\alpha$ stands for protons  and alpha particles respectively).
In Eq. (\ref{eq:Vlasov}), $f_i({\bf r}, {\bf v}, t)$ is the ion distribution function, ${\bf E}({\bf r},t)$ and ${\bf B}({\bf r},t)$ are the electric and magnetic fields. In Eq. (\ref{eq:ohm}), the electron bulk velocity is defined as ${\bf u}_e=(\sum_i Z_i n_i {\bf u}_i- {\bf \nabla} \times {\bf B})/n_e$, where the ion bulk velocities ${\bf u}_i$ are evaluated as first order velocity moments of the ion distribution function; finally, ${\bf j}=\nabla\times {\bf B}$ represents the total current density. 
An isothermal equation of state for the electron pressure $P_e$ closes the system.

In Eqs. (\ref{eq:Vlasov})-(\ref{eq:farampere}) time is scaled by the inverse proton-cyclotron frequency $\Omega_{cp}^{-1}$, velocities by the Alfv{\'e}n speed $V_{A}$, lengths by the proton skin depth $d_{p}=V_{A}/\Omega_{cp}$ and masses by the proton mass $m_p$. From now on, all the physical quantities will be expressed in units of the characteristic parameters listed above. A small value for the resistivity $(\eta = 2\times10^{-2})$ has been chosen in order to achieve relatively high Reynolds numbers and to remove any spurious numerical effects due to the presence of strong current sheets. The initial equilibrium consists of a plasma composed of kinetic protons and alpha particles, with Maxwellian velocity distributions and homogeneous densities ($n_{0,p}$ and $n_{0,\alpha}$ respectively), and fluid electrons embedded in a background magnetic field ${\bf B}_0 = B_0{\bf e}_z$. The plasma dynamics and the development of turbulence are  investigated in a double periodic $(x,y)$ domain perpendicular to ${\bf B}_0$, where the total magnetic field can be written as ${\bf B}={\bf B}_0 + {\bf B}_\perp$. 

\begin{figure}
\begin{center}
\includegraphics [width=6.2cm]{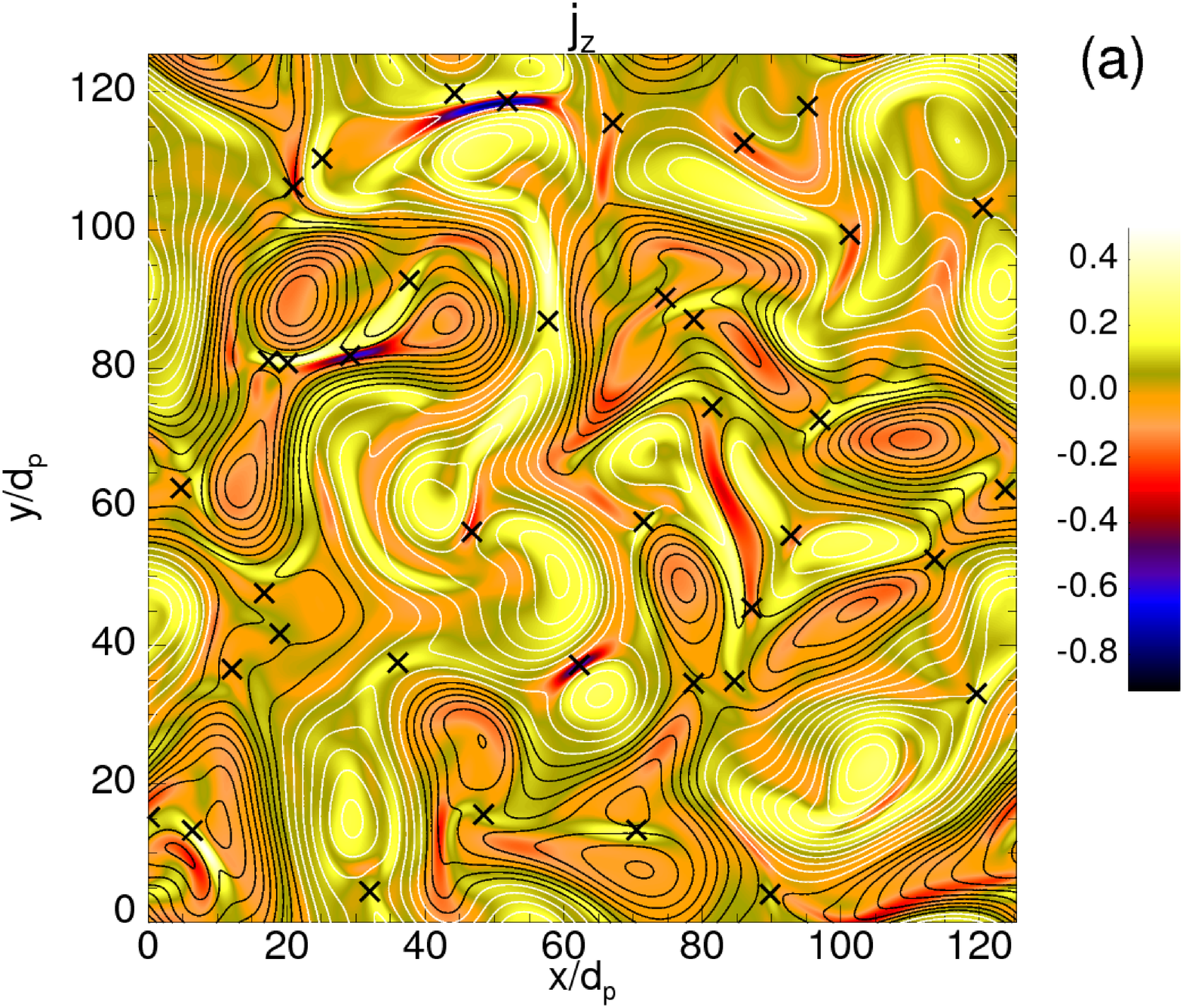}
\includegraphics [width=6.2cm]{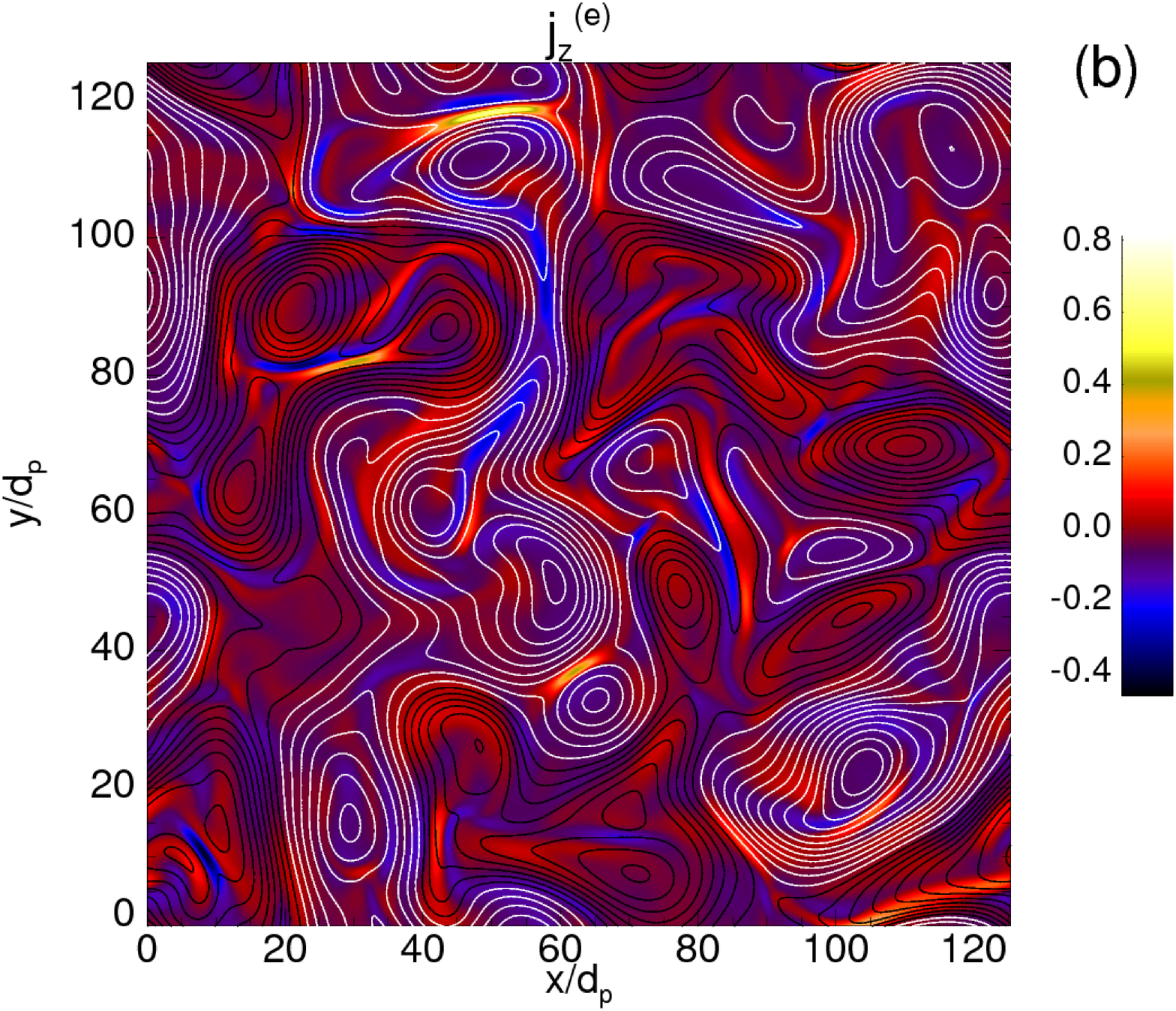}
\includegraphics [width=6.2cm]{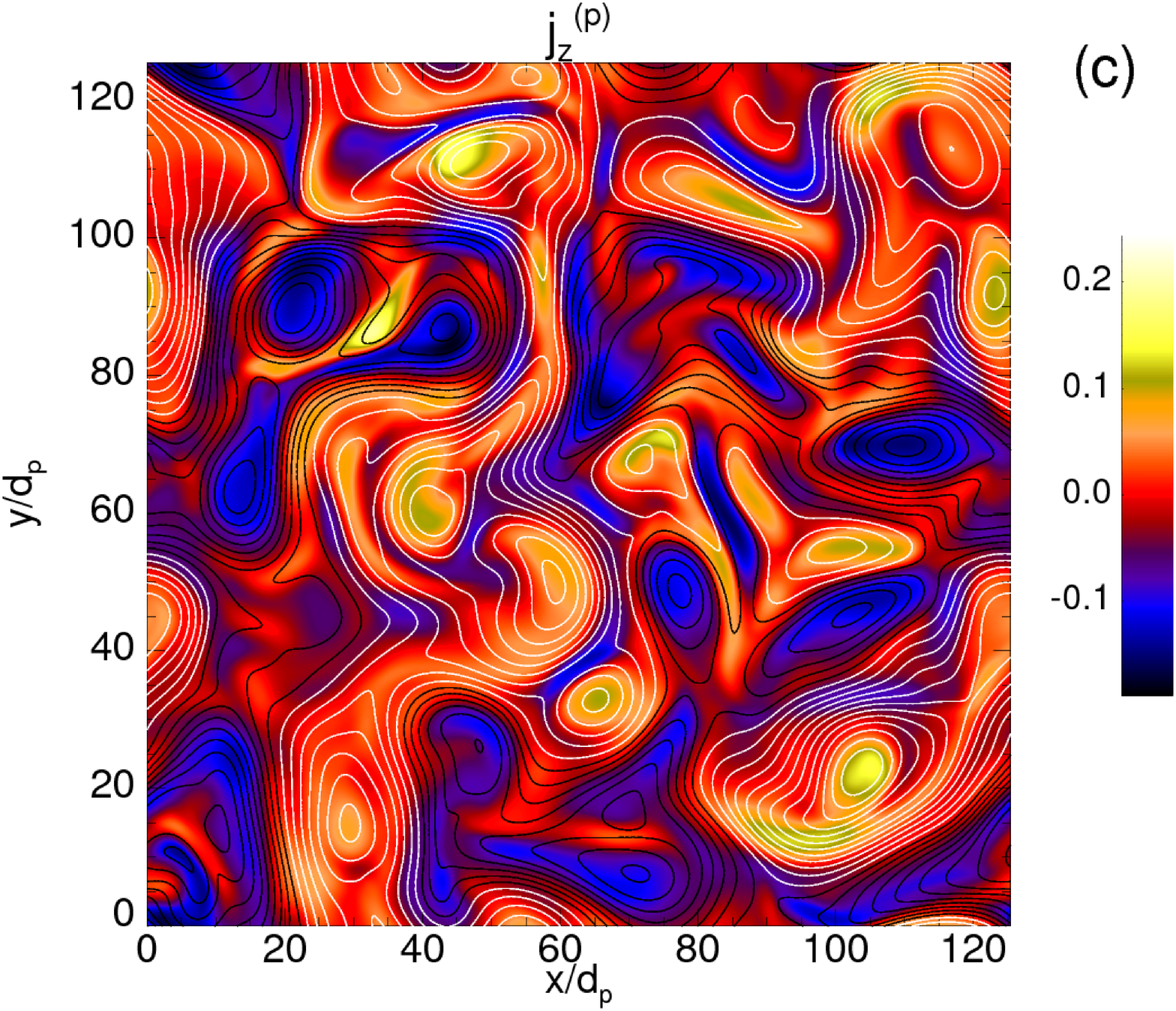}
\includegraphics [width=6.2cm]{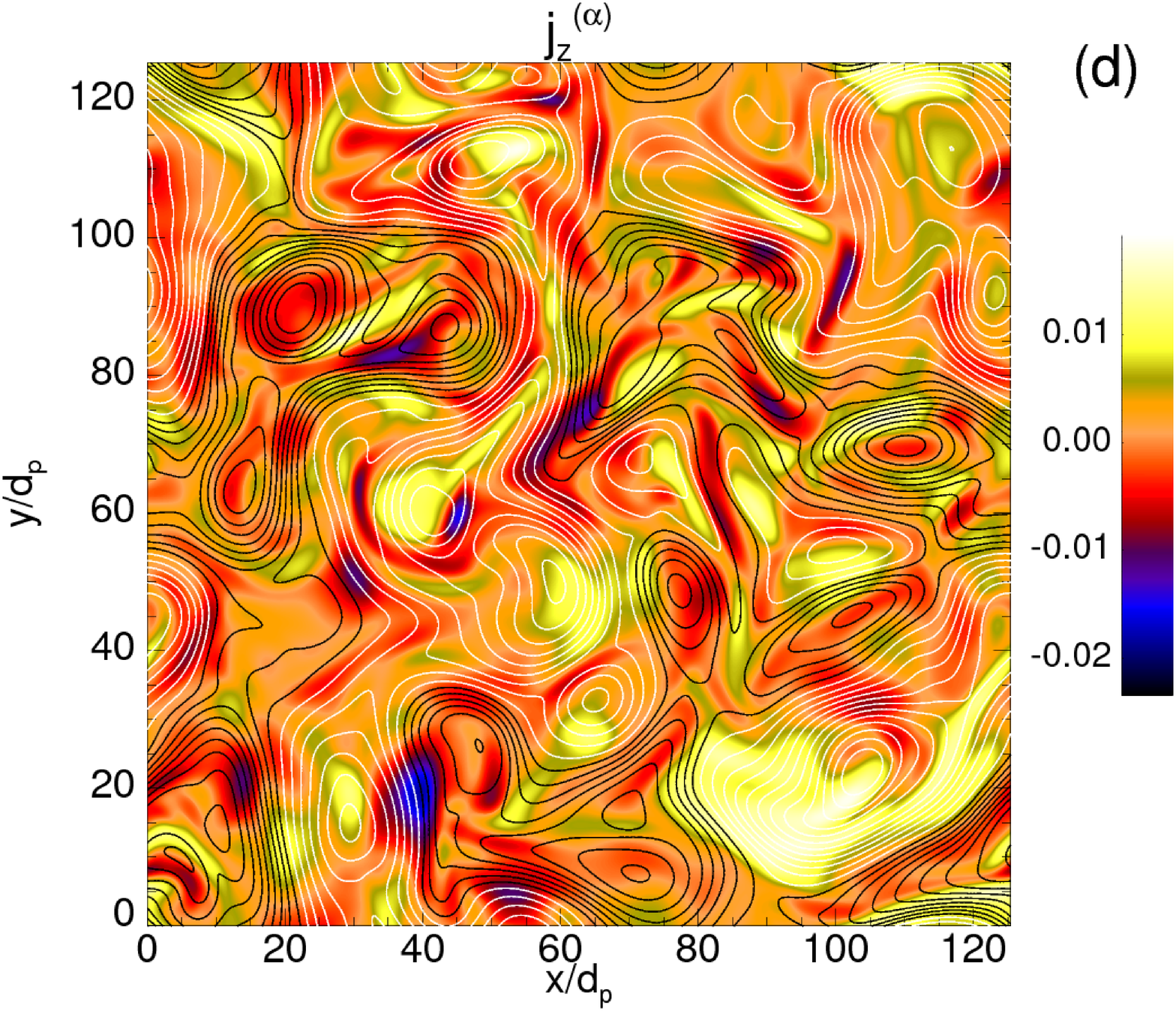}
\caption{Contour plots (shaded colors) of the out-of plane current densities: $j_z$ (a), $j_z^{(e)}$ (b), $j_z^{(p)}$ (c), and $j_z^{(\alpha)}$ (d). The isolines of the magnetic potential $A_z$ are indicated by black/white lines. In panel (a), the positions of the X-points (regions of magnetic reconnection) are indicated by black thick crosses.}
\label{fg:jz}
\end{center}
\end{figure}

The equilibrium configuration is perturbed by a 2D spectrum of fluctuations for
the magnetic and proton velocity fields (alpha particles have zero initial bulk velocity).
We inject energy with random phases and wave numbers
in the range $0.066 < k < 2$, where $k=2\pi m/L$, with $2 \leq m \leq
6$ and $L$ being the box size in each spatial direction. The rms of the initial magnetic perturbations is $\delta B/B_0\simeq 0.3$. Neither density
disturbances nor parallel variances are imposed at $t=0$.
%
%
The proton plasma beta is $\beta_p=2{v_{th,p}}^2/V_{A}^2=2$ (where $v_{th,p} = \sqrt{T_p/m_p} = 1$ is the
proton thermal speed) and the electron to proton temperature ratio is $T_e/T_p=1$.
For the alpha particles we set $Z_{\alpha}=2$; 
$m_{\alpha}/m_p=4$, $n_{0,\alpha}/n_{0,p}=5\%$ and $T_{\alpha}/T_p=1$. With this choice, the alpha particle thermal speed is $v_{th,\alpha}=v_{th,p}/2$.

The system size in the spatial domain is $L=2\pi \times 20 d_p$ in both $x$ and $y$ directions, while
the limits of the velocity domain for both ion species are fixed at
$v_{max,i}=\pm 5 v_{th,i}$ in each velocity direction. In
these simulations, we use $512^2$ grid-points in the two-dimensional spatial domain 
and $61^3$ and $31^3$ grid-points in proton and alpha particle three-dimensional velocity domains, respectively.
We point out that in the Ohm equation for the electric field we have neglected the electron inertia terms. These terms are in fact proportional to the squared electron skin depth (which in scaled units is given by $d_e^2=m_e/m_p$), then cannot be adequately resolved within the discretization of our simulations.
The time step $\Delta t$ has been chosen in such a way that the Courant-Friedrichs-Lewy condition for the numerical stability of the Vlasov algorithm is satisfied \cite[]{pey86}.

\section{Numerical Results}


\begin{figure}
\begin{center}
\includegraphics [width=7cm]{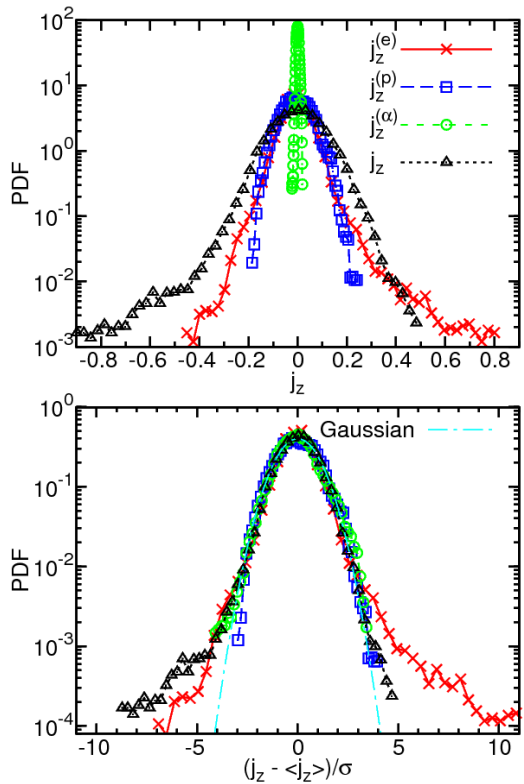}
\caption{In the top panel we show the PDFs of the different current densities: $j_z^{(e)}$ red solid-cross line, $j_z^{(p)}$ blue dashed-square line, $j_z^{(\alpha)}$ green dashed-circle line, and $j_z$ black dashed-triangle line. In the bottom panel, the PDFs of the standardized variables, obtained by subtracting the average and normalizing to the respective rms value, are reported. The light-blue dot-dashed line is the Gaussian fit.}
\label{fg:pdfjz}
\end{center}
\end{figure}

We numerically study the kinetic evolution of protons and alpha particles in a situation of decaying turbulence.
We expect that kinetic effects develop simultaneously with magnetic fluctuations and shears, the latter playing a fundamental role in the production of interesting features such as particle acceleration, heating, temperature anisotropy, wave-particle like interactions, and generation of beams in the ion distribution function.

As in the fluid counterpart, large scale fluctuations produce a turbulent cascade toward 
small scales (high $k$'s). In analogy with fluid models (MHD, Hall MHD, etc.) of decaying turbulence \cite[]{min09}, 
it is possible to identify an instant of time at which the turbulent activity reaches its maximum value.
Since the current density is proportional to the level of small-scale gradients,
a good indicator of the level of turbulent activity is represented by the average out-of-plane squared
current density $\langle j_z^2\rangle$, whose time evolution is shown in Figure \ref{fg:turbulence}.
Evidently, at $t = t^* \sim 40$ $\langle j_z^2\rangle$ attains its maximum value. This is the characteristic time at which decaying turbulence shares many statistical similarities with steady state (driven) turbulence, and, at this time, we perform our analysis.

\begin{figure}
\begin{center}
\includegraphics [width=8.6cm]{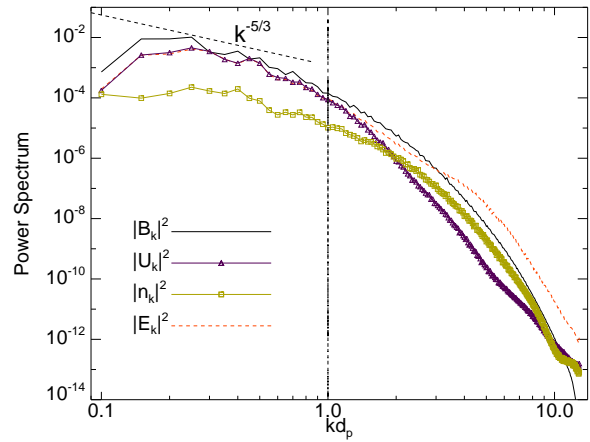}
\caption{Power spectra of $n_p$ (green-square line), ${\bf u}_p$ (purple-triangle line), ${\bf B}$ (black-solid line) and ${\bf E}$ (red-dashed line).}
\label{fg:spectrum}
\end{center}
\end{figure}

The turbulent activity leads to the generation of coherent structures. Vortices and current sheets appear in the contour plots of Figure \ref{fg:jz}, where shaded contours of the out-of-plane total current density $j_z$ [panel (a)] are represented. In the same figure, for the descending panels, the separate contributions $j_z^{(i)}$ of each species, namely $j_z^{e}$ (electron current), $j_z^{p}$ (proton current), and $j_z^{\alpha}$ (alpha-particle current) are also shown. The contour lines in each plot represent the magnetic potential $A_z$ of the inplane magnetic field (${\bf B}_{\perp} = \nabla A_z \times {\bf e}_z$). The different colors (black/white) of the $A_z$ contour lines indicate different directions of rotation of the vortices. At a careful analysis of the four plots of Figure \ref{fg:jz}, a certain correlation seems to exist between the proton and the alpha-particle current densities, revealing that the local small-scale structures of the two ion species behave in a similar way. Moreover, the electron flow generates structures at smaller scales, manifesting a more intermittent and bursty character. The coherent structures visible in the four plots of Figure \ref{fg:jz} are not static, but evolve in time interacting nonlinearly among each others. Moreover, in between the islands, the current becomes very intense and magnetic reconnection events locally occur at the $X$ points of $A_z$, indicated in the contour plot of panel (a) by black crosses. The presence of these high magnetic stress regions is a signature of the intermittent nature of the magnetic field, and this affects also the patchiness of the parallel and perpendicular heating (see below).

In the top panel of Figure \ref{fg:pdfjz} we report the probability distribution functions (PDF) of $j_z^{(e)}$ (red solid-cross line), $j_z^{(p)}$ (blue dashed-square line), $j_z^{(\alpha)}$ (green dashed-circle line), and $j_z$ (black dashed-triangle line). This plot clearly indicates that there is a certain ordering in the maximum values of the achieved current. The main contribution to the total current seems to come from the electrons and the protons, that develop the most intense bursty events. In contrast, the alpha-particle current structures are smoother and are concentrated on larger scales, as can be noticed from Figure \ref{fg:jz}.  In the same figure (bottom panel), we report the PDF of the standardized variables obtained by subtracting the average and normalizing to the respective rms value. The Gaussian fit is also plotted (light-blue dot-dashed line) as reference. The currents $j_z$ and $j_z^{(e)}$ are  highly non-Gaussian distributed, because they are related to the increments (gradients) of the magnetic field (and electron flows are essentially frozen-in). The proton and alpha particle contributions, on the other hand, behave more like Gaussian variables, since they are related to primitive variables of turbulence such as velocities and densities, and they do not capture high order statistics.

\begin{figure}
\begin{center}
\includegraphics [width=8.4cm]{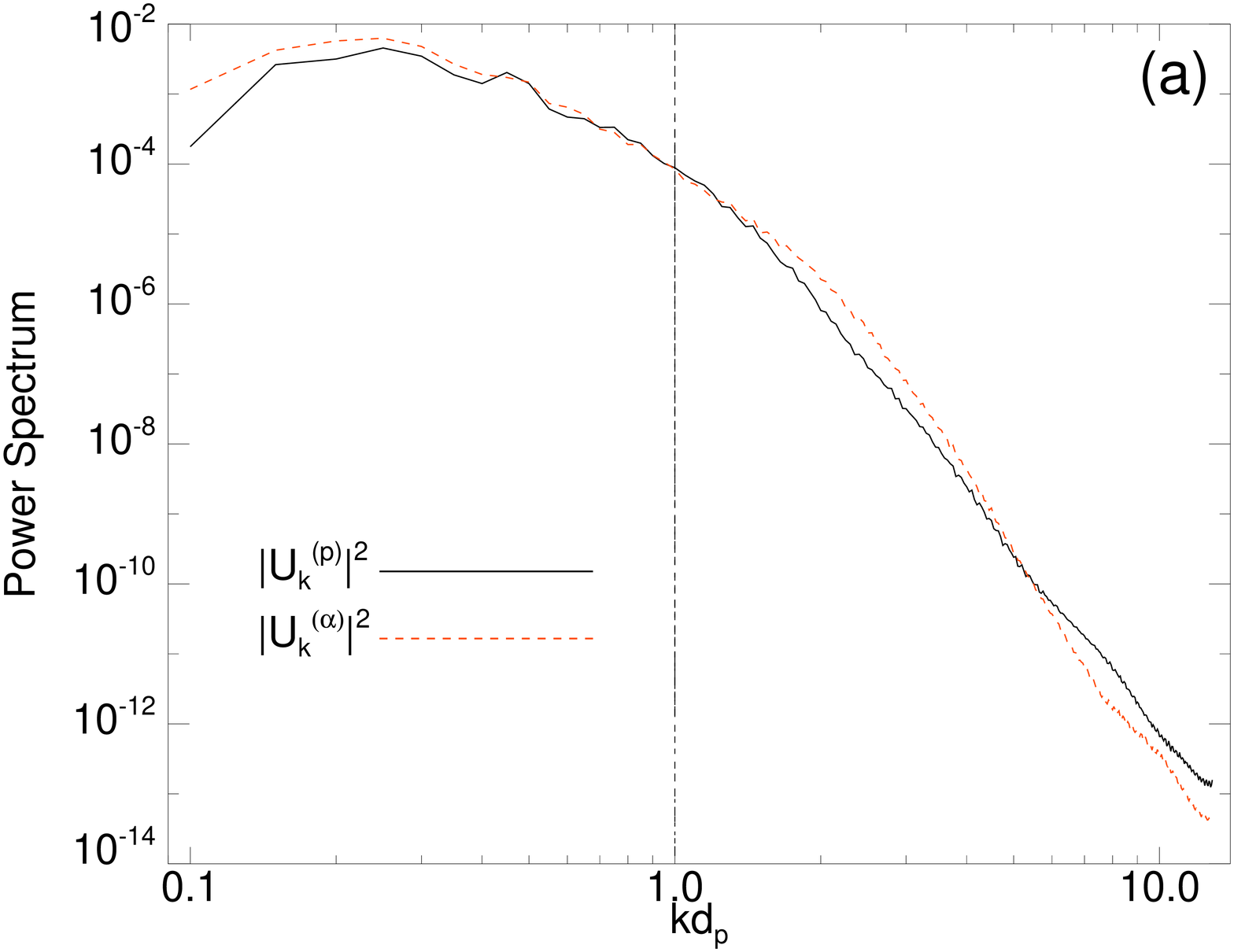}
\includegraphics [width=8.4cm]{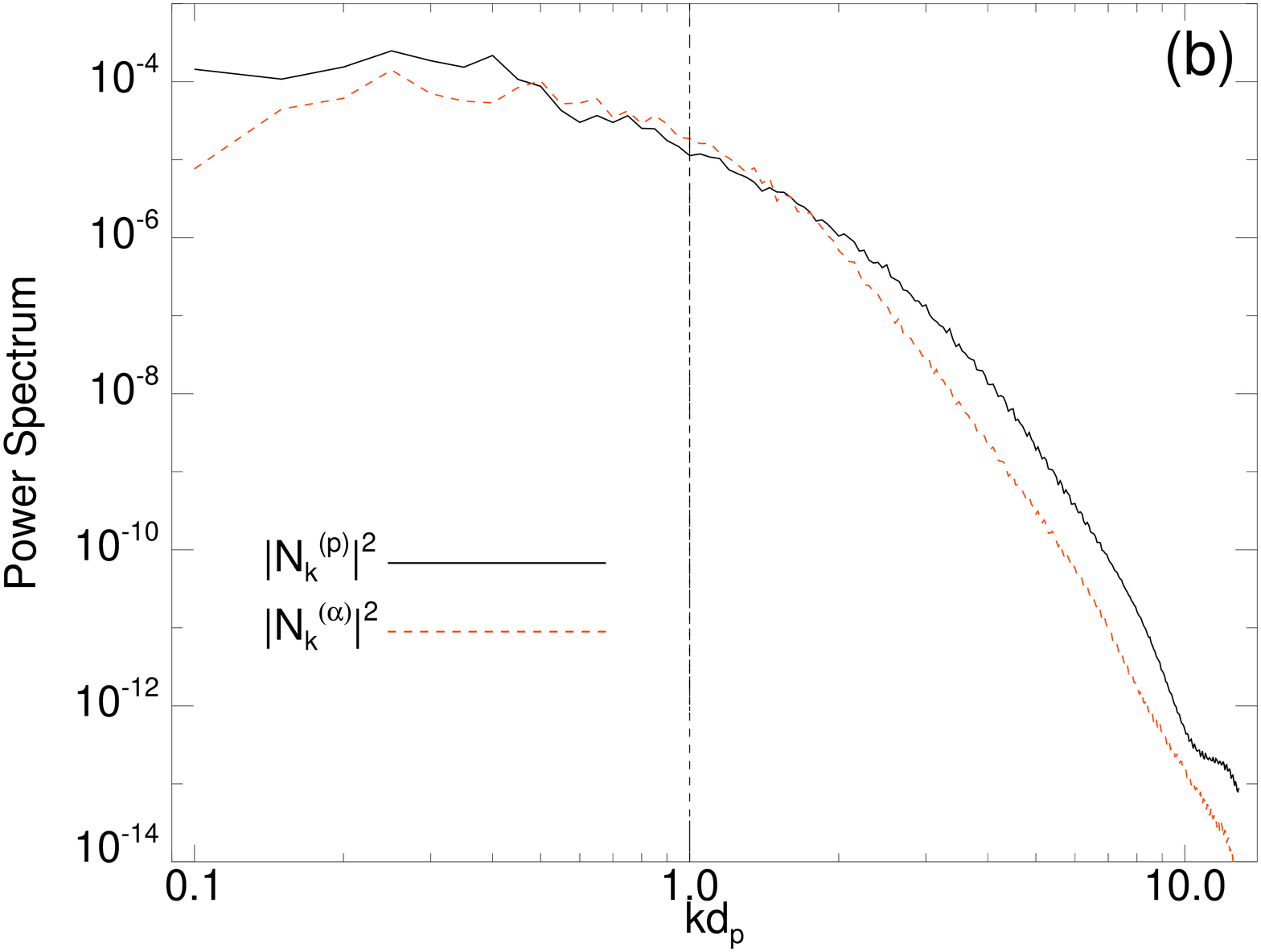}
\caption{Panel (a): power spectra $|U_k^{(i)}|^2$ of the proton bulk velocity ($i=p$, black-solid line) and of the alpha-particle bulk velocity ($i=\alpha$, red-dashed line); panel (b): power spectra $|N_k^{(i)}|^2$ of the proton density ($i=p$, black-solid line) and of the alpha-particle density ($i=\alpha$, red-dashed line)}.
\label{fg:comparison}
\end{center}
\end{figure}

In order to quantify the turbulence, the power spectra of density and bulk velocity for both protons and alpha particles and of magnetic and electric fields have been computed. In Figure \ref{fg:spectrum} we show the power spectra of $n_p$ (green-square line), ${\bf u}_p$ (purple-triangle line), ${\bf B}$ (black-solid line) and ${\bf E}$ (red-dashed line). The Kolmogorov expectation $k^{-5/3}$ (black-dashed line) has been plotted in Figure \ref{fg:spectrum} as a reference.
These omnidirectional power spectra reveal several interesting features, many of them recovered also in solar-wind spacecraft observations. In fact, the large scale activity is incompressible and the Alfv\`{e}nic correlation between magnetic and velocity fluctuations is broken at the proton skin depth (vertical black dashed line). Moreover, the electric activity (red-dashed line) at higher wavenumbers is significantly more intense than the magnetic one (black-solid line) \cite[]{bru05, bal05}. It is worth to point out that the power spectra displayed in Figure \ref{fg:spectrum} present no significant differences with respect to the same spectra obtained through HVM simulations without alpha particles [see Figure 1(b) in \cite{ser12}], meaning that the presence of alpha particles does not significantly affect the dynamical evolution of the turbulent cascade.
To make a direct comparison of the dynamical evolution of the two ion species, in Figure
\ref{fg:comparison} we show the velocity spectra $|U_k^{(i)}|^2$ [panel (a)] and the normalized density spectra $|N_k^{(i)}|^2$ [panel (b)]
for protons ($i=p$, black-solid lines) and for alpha particle ($i=\alpha$, red-dashed lines).
The density spectra for protons and alpha particles are normalized to $n_{0,p}$ and $n_{0,\alpha}$ respectively.
While the velocity spectra [panel (a)] of the two ion species do not display significantly different features,
we notice that the alpha particles contribution to the density spectra [panel (b)] is lower than the proton one
for wavenumbers higher than the proton skin depth wavenumber. This
behavior is possibly related to the fact that the alpha particles are heavier than
protons, so their inertia does not allow to follow the field fluctuations at smaller scales.

\begin{figure}
\begin{center}
\includegraphics [width=9.2cm]{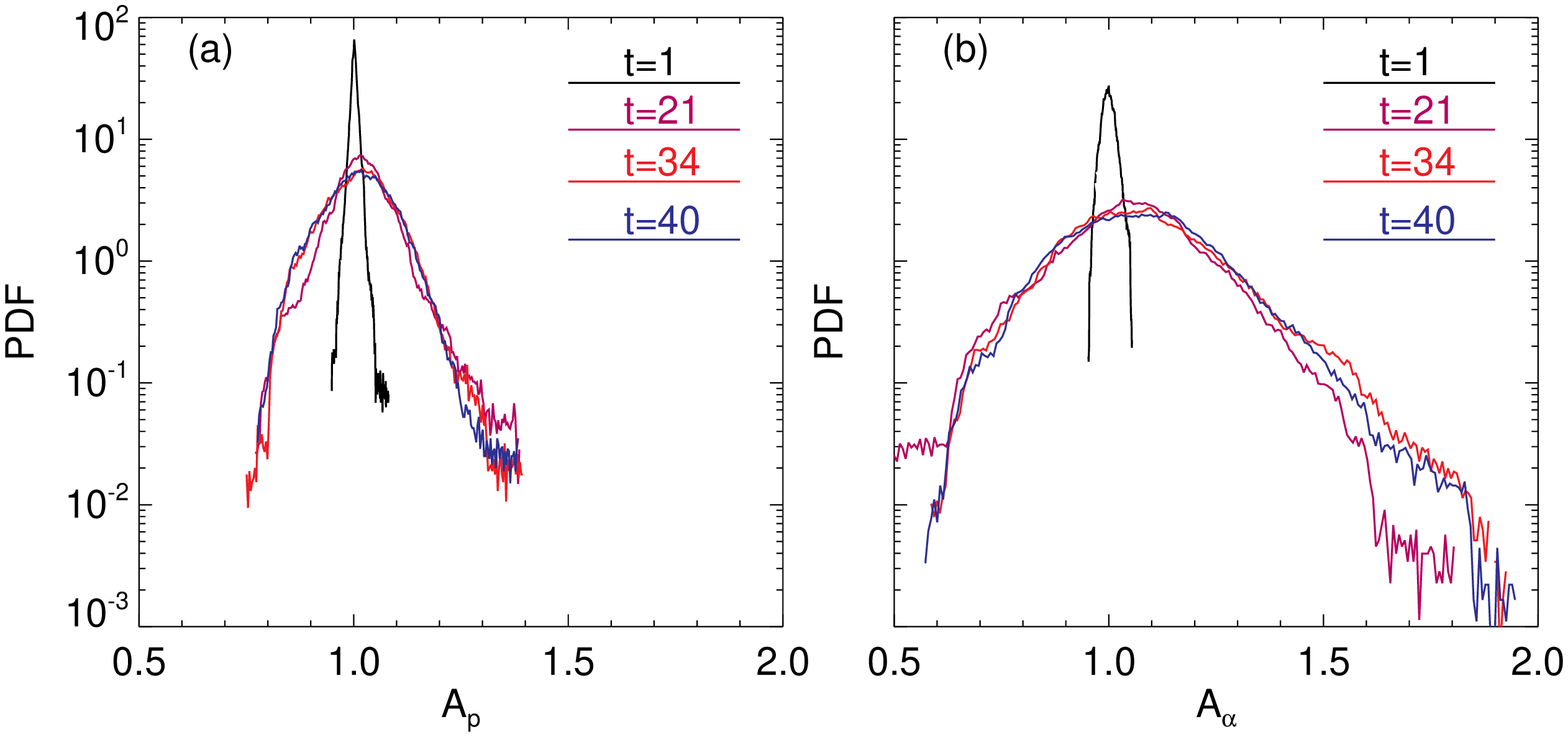}
\includegraphics [width=7.2cm]{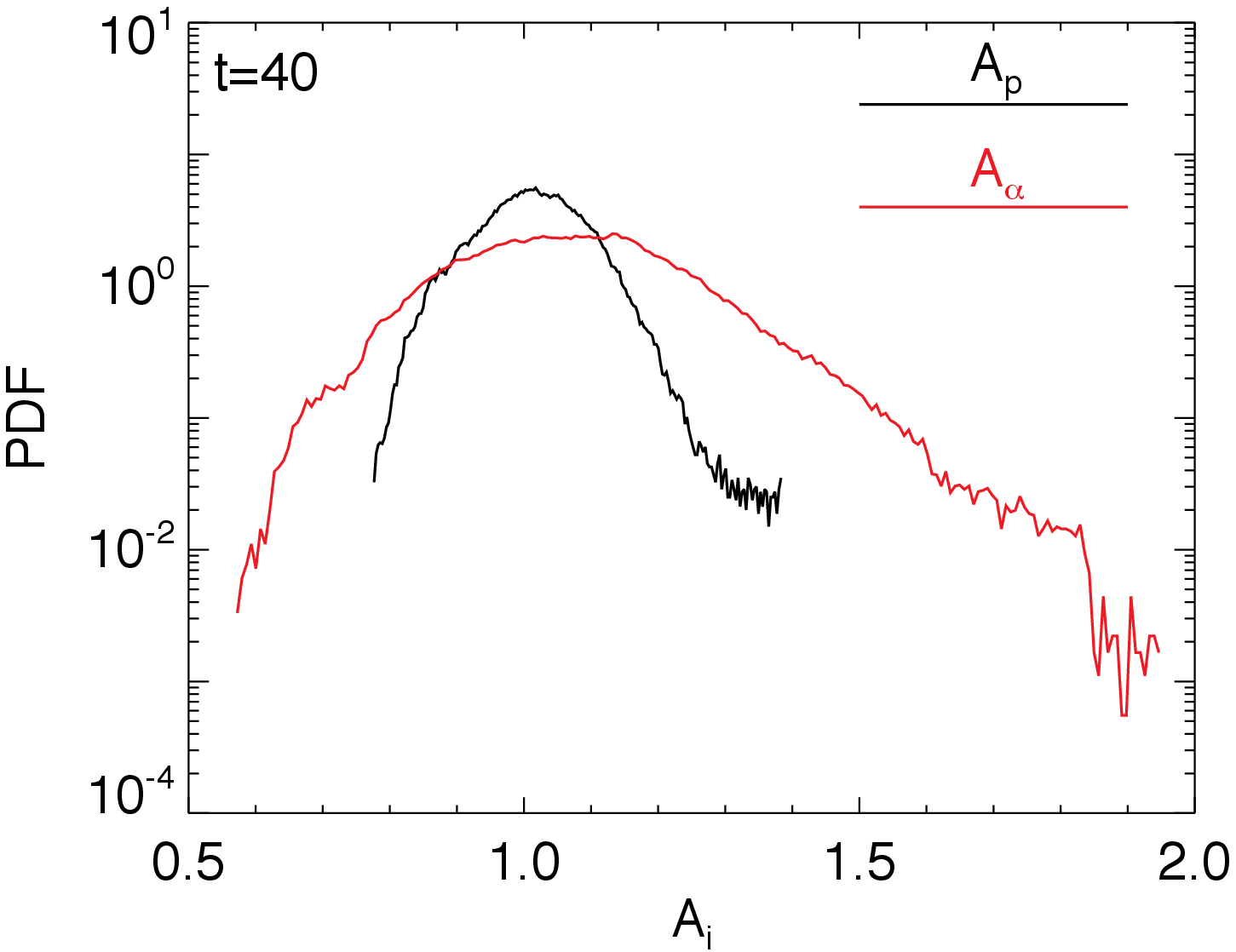}
\caption{Top panel: PDF of the temperature  anisotropy $A_i= T_{\perp}^{(i)}/T_{\parallel}^{(i)}$ of protons (a) and alpha-particles (b) at four different times in the system evolution: $t=1$ black line, $t=21$ purple line, $t=34$ red line, $t=40$ blue line.
Bottom panel: comparison between the proton ($A_p$, black line) and the alpha-particle ($A_{\alpha}$, red line) anisotropies is reported at the peak of the turbulent activity, i.e., at $t=40$.}
\label{fg:pdf}
\end{center}
\end{figure}

At this point it is important to investigate the link between the turbulent
behavior observed in the plasma and the generation of non-Maxwellian features in 
velocity space. For this purpose, the ion temperature anisotropy for each species $A_i=T_{\perp}^{(i)}/T_{\parallel}^{(i)}$, defined
as the ratio between the perpendicular and the parallel temperature with respect to the local 
magnetic field, has been computed. Our initial condition has been set up in such a way to have spatially isotropic temperatures for both the ion species at $t=0$. Nevertheless, during the development of turbulence the temperatures do not remain spatially isotropic but present local enhancements and depressions nearby the regions of high magnetic stress (not shown), as already found for the protons in \cite{ser12}.

Figure \ref{fg:pdf} shows the probability distribution function (PDF) of the temperature
anisotropy for protons $A_p$ [panel (a)] and alpha particles $A_\alpha$ [panel (b)] at four different times in the simulation.
In the early stage of the system evolution ($t=1$ black line), the PDFs are picked around $A_p=A_\alpha=1$, meaning that the simulation starts with an isotropic configuration. During the evolution of the system ($t=21$, purple line; $t=34$, red line; $t=40$, blue line) the PDFs elongate in the parallel $(A_i<1)$ and in the perpendicular $(A_i>1)$ direction, displaying a strong anisotropic behavior. It is worth noting that the statistical behavior of the anisotropies saturates already at $t\simeq 20 \Omega_{cp}^{-1}$.
Regardless of the particular ion species, the anisotropy preferentially manifests itself along the perpendicular direction,
an evidence commonly detected in the solar-wind observations \cite[]{bou10, bou11_b}.
However, alpha particles are more anisotropic than protons, as more evidently shown in Figure \ref{fg:pdf} (bottom plot) where we have directly compared the PDFs of the two ion species ($A_p$ black line and $A_\alpha$ red line) when the peak of the nonlinear activity is reached ($t=40\Omega_{cp}^{-1}$).

\begin{figure}
\begin{center}
\includegraphics [width=7.5cm]{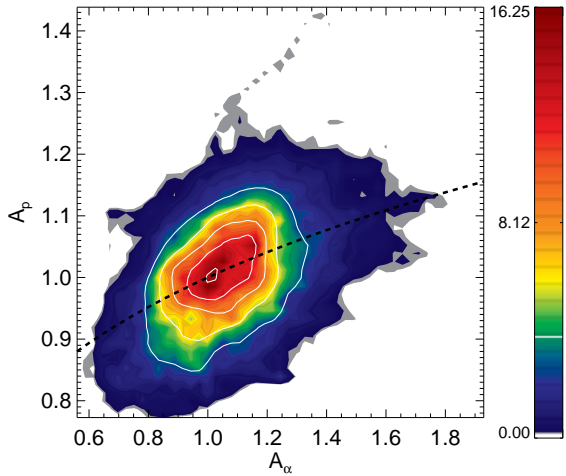}
\caption{Joint probability distribution function of proton and alpha-particle temperature anisotropy. This two-dimensional PDF shows a correlation between the anisotropy in the two species. The best fit $A_p\sim A_\alpha^\zeta$, with $\zeta=0.22$, is also reported with a thick solid line. This behavior is in good agreement with solar wind data, as it can be seen comparing this plot with Figure 1-(b) in \cite{maruca12}.}
\label{fg:2d}
\end{center}
\end{figure}

A question now naturally occurs: are these patchy anisotropies correlated? Any correlation between $A_\alpha$ and $A_p$ may reveal that simultaneous kinetic instabilities locally occur, modulated by the ambient magnetic field, or that an instability for a given species may influence the other, and vice-versa.
In Figure \ref{fg:2d} we analyze the correlation between protons and alpha particle temperature anisotropy, showing the joint PDF. 
Although most of the events are concentrated at $A_p=A_{\alpha}=1$ (isotropy), and are broadly scattered because of turbulence, this joint distribution suggests that there is a clear monotonic dependency between alpha and proton anisotropies. The shape of this distribution is in good agreement with solar wind data, as it can be seen comparing our Figure \ref{fg:2d} with Figure 1 of \cite{maruca12}. Moreover, analogously to \cite{maruca12}, we fitted the above distribution with $A_p=A_\alpha^\zeta$, obtaining $\zeta\simeq0.22$ (note that in \cite{maruca12} the authors obtained $\zeta\simeq0.25$). These results suggest that the correlation between proton and alpha particle kinetic effects, commonly observed in the solar wind, may be the result of an active turbulent cascade, where kinetic instabilities are locally activated and modulated by the ambient magnetic field.

\begin{figure}
\begin{center}
\includegraphics [width=7.5cm]{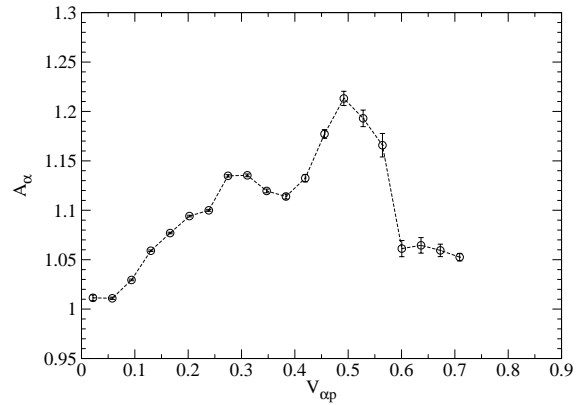}
\caption{Anisotropy of the alpha-particle temperatures binned as a function of the differential speed $V_{\alpha p}=|{\bf u}_p-{\bf u}_\alpha|$ (in Alfv\'enic units). This behavior is in good agreement with solar wind observations (cfr. with Figure 4 of \cite[]{bou11_b}).}
\label{fg:2d2}
\end{center}
\end{figure}

In a multi-ion plasma another source of instability is represented by the differential flow between different ion species \cite[]{gar93,gar06,gar08}.
We found that the temperature anisotropy for the alpha particles shows a certain correlation to the their drift velocity $V_{\alpha p}=|{\bf u}_p-{\bf u}_{\alpha}|$ with respect to protons. This is visible in Figure \ref{fg:2d2}, where we report $A_\alpha$ as a function of $V_{\alpha p}$: the temperature anisotropy increases with increasing relative flow speeds (in Alfvenic units), up to $V_{\alpha p}\sim0.5$.

These results are again in good agreement with some observational analyses. \cite{bou11_b} studied correlations of temperature anisotropies and differential ion speed in the solar-wind measurements from the Helios spacecraft; for the case of the alpha particles, they found that $A_\alpha$ increases as the ion differential speed stays below about $0.5 V_{A}$. Beyond this value $A_\alpha$ becomes roughly constant, until $V_{\alpha p}$ exceeds a value of about $0.7 V_A$, but then it decreases towards a value below unity when $V_{\alpha p}\simeq V_{A}$ (not reached in our system).  Comparing our Figure \ref{fg:2d2} with Figure 4 of \cite{bou11_b}, we find a very good correspondence. However, it is worth to point out that in a different data analysis of Advanced Composition Explorer (ACE) solar wind observations, \cite[]{kas08} found that the alpha temperature anisotropy is monotonically decreasing with increasing alpha particle to proton relative speed in the range $0\leq V_{\alpha p}\lesssim 0.5$. It is also worth noting that while these studies are carried out on years of solar wind data, that detect different plasmas with different parameters, homogeneities, large scale effects and so on, in our case, these phenomena are the genuine result of a turbulent and statistically homogeneous cascade.

\begin{figure}
\begin{center}
\includegraphics [width=8.5cm]{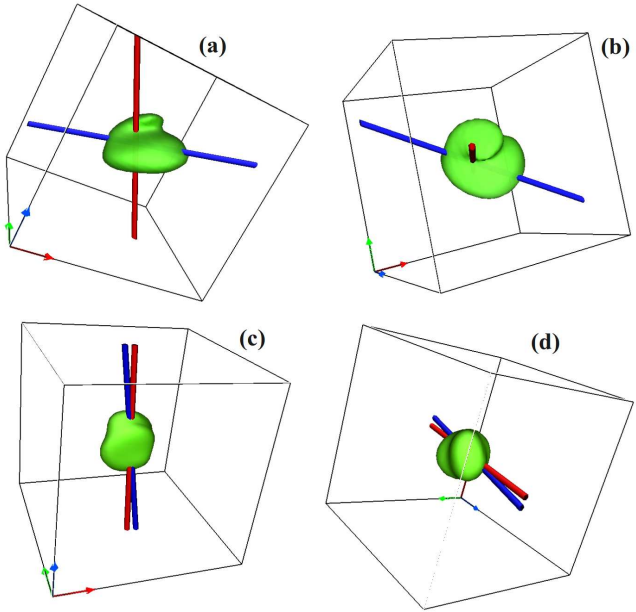}
\caption{Iso-surfaces of the alpha particle velocity distribution $f_\alpha({\bf r}^*,v_x,v_y,v_z)$, at two different spatial locations: namely in regions where the distribution function shows anisotropy $A_\alpha>$1 (a)-(b) and $A_\alpha<$1 (c)-(d). The direction of the local magnetic field (red tube) is also reported together with the principal axis (blue tube) of the velocity distribution (see \cite{ser12}).}
\label{3Dalpha}
\end{center}
\end{figure}

To conclude our study, we discuss a few examples of the effects of turbulence on the velocity distributions of alpha particles. In Figure \ref{3Dalpha}, we show the iso-surfaces of the alpha-particle velocity distribution at two distinct locations in physical space, at which $A_\alpha>1$ [(a)-(b)], and $A_\alpha<1$ [(c)-(d)]. In the same figure, we also report the direction of the local magnetic field (red tube) and the principal axis (blue tube) of the velocity distribution, evaluated from the stress tensor in the minimum variance frame (for more details see \cite{ser12}). The alpha-particle velocity distribution appears strongly affected by turbulence and modulated by the local magnetic field topology, manifesting both kinds of anisotropy; moreover, as it is clear from Figure \ref{3Dalpha}, the principal axis of the velocity distribution can be both aligned or perpendicular to the local magnetic field. Another interesting feature that can be appreciated in the plots of Figure \ref{3Dalpha} is the local formation of bubbles in the velocity distribution along the direction of the local magnetic field that resemble the characteristic longitudinal beams of accelerated particles commonly observed in the solar wind data \cite[]{mar82_a, mar82_b} and in 1D-3V HVM simulations \cite[]{val08,per11}. In these numerical papers a possible mechanism responsible for the generation of these beams of accelerated particles has been identified in the resonant interaction of particles with a newly identified branch of electrostatic fluctuations called ion-bulk waves \cite[]{val11}. This resonant interaction results more efficient for protons than for alpha particles \cite[]{per11}, in agreement with recent solar-wind data analysis \cite[]{mar10}.

\section{Summary and Discussion}

In this paper we discussed the numerical results of Hybrid Vlasov-Maxwell simulations employed to investigate the role of kinetic effects in a two-dimensional turbulent multi-ion plasma, composed of protons, alpha particles and fluid electrons, in the typical conditions of the solar-wind environment. While we pointed out that the presence of a small percentage of heavy ions does not affect the evolution of the turbulent cascade, our numerical results clearly show that the dynamics of alpha particles at short spatial scales displays several interesting aspects, mainly consisting in the departure of the distribution function from the typical Maxwellian configuration.

In a situation of decaying turbulence, coherent structures appear, such as vortices and current sheets. In between magnetic islands, reconnection events occur. We quantify the contribution to the total current due to protons, alpha particles and electrons, finding that the most of the contribution comes from the electrons, while the current density of alpha particles is very low, due to their higher inertia. Moreover, we found the currents of protons and alpha particles are Gaussian distributed, while electron current densities are highly intermittent, since they capture the small-scale features of the magnetic field.

Power spectra for protons and for the electromagnetic fields are not too much affected by the presence of alpha particles, while at smaller scales (at k's higher than the inverse proton skin depth), the power spectrum of the alpha density has a steeper slope than that of proton density.

The non-Maxwellian features generated during the evolution of turbulence have been quantified through the statistical analysis of the temperature anisotropy in the reference frame of the local magnetic field. The joint probability distribution of the temperature anisotropies of alpha particles and protons suggests that there is a clear monotonic dependency between alpha and proton anisotropies. This reveals the occurrence of simultaneous local kinetic instabilities or the influence of the dynamical evolution of one species on the other. In general, for the parameters studied here, we found that alpha particles develop higher anisotropy than protons.

With the purpose of analyzing the role of kinetic instabilities driven by the relative speed of the two ion species, we evaluated the dependence of the temperature anisotropy of alpha particles on their drift speed $V_{\alpha p}$ with respect to protons. From this analysis, we found that the temperature anisotropy increases with increasing relative flow speeds up to $V_{\alpha p} \simeq 0.5$ (in units of $V_A$). Moreover, the velocity distributions of alpha particles create bumps along the local magnetic field, resembling very common structures observed in the solar wind \cite{mar06}.

By comparing our numerical results to recent solar-wind analyses \citep{maruca12,bou11_b}, we found a very good quantitative correspondence both for the correlation of alpha particle and proton temperature anisotropies and for the correlation of alpha anisotropy and relative flow speed. However, it is worth noting that, while the observational studies are carried out on years of solar wind data, that detect plasmas with very different features and in different physical regimes, in our case these correspondences are the genuine results of a turbulent cascade, where locally both the magnetic field topology and the relative motion of different ion flows can be the main sources of kinetic effects.

The results presented in this work significantly reproduce an important part of the complex phenomenology underlying many processes in the solar wind and suggest that a noise-free Eulerian Vlasov description of a multi-component collision-free plasma plays a fundamental role in the interpretation of the observational data from spacecraft.

\section*{acknowledgments}

The numerical simulations discussed in the present paper were performed within the project ASWTURB 2011 (HP10BO2REM), supported by the Italian SuperComputing Resource
Allocation, ISCRA-CINECA, Bologna, Italy. S. S. acknowledges the Marie Curie Project FP7 PIRSES-2010-269297 - ``Turboplasmas'', and the POR Calabria FSE 2007/2013.


\begin{thebibliography}{}

\bibitem[Alexandrova et al.(2009)]{ale09} Alexandrova, O., Saur, J., Lacombe,
C, Mangeney, A., Mitchell, J., Schwartz, S., J., \& Robert, P. 2009, Phys. Rev.
Lett., 103, 165003

\bibitem[Araneda et al.(2008)]{ara08} Araneda, J. A., Marsch, E., \&  Vi\~{n}as,
A. F. 2008, Phys. Rev. Lett., 100, 125003

\bibitem[Araneda et al.(2009)]{ara09} Araneda, J. A., Maneva, Y., \& Marsch, E.
2009, Phys. Rev. Lett., 102, 175001

\bibitem[Bale et al.(2005)]{bal05} Bale, S. D., Kellogg, P. J., Mozer,
F. S., Horbury, T. S., \& Reme, H. 2005, Phys. Rev. Lett., 94, 215002

\bibitem[Bourouaine et al.(2010)]{bou10} Bourouaine, S., Marsch, E., \& 
Neubauer, F. M. 2010, Geophys. Res. Lett., 37, L14104

\bibitem[Bourouaine et al.(2011a)]{bou11_a} Bourouaine, S., Marsch, E., \& 
Neubauer, F. M. 2011a, ApJ, 728, L3

\bibitem[Bourouaine et al.(2011b)]{bou11_b} Bourouaine, S., Marsch, E., \&
Neubauer, F. M. 2011b, A\&A, 536, A39

\bibitem[Bourouaine et al (2012)]{bourouaine12} Bourouaine S., Alexandrova O., Marsch E., \&
Maksimovic M. 2012, ApJ, 749, 102.

\bibitem[Bruno \& Carbone(2005)]{bru05} Bruno, R., \& Carbone, V. 2005, Living
Rev. Solar Phys., 2, 4

\bibitem[Camporeale \& Burgess(2011)]{cam11} Camporeale, E., \& Burgess, D. 2011,
ApJ, 730, 114

\bibitem[Drake et al.(2010)]{dra10} Drake, J. F., Opher, M., Swisdak, M., \&
Chamoun, J. N. 2010, ApJ, 709, 963

\bibitem[Gary(1993)]{gar93} Gary, S. P., 1993, Theory of Space Plasma Microinstabilities,
Cambridge UUnievrsity Press

\bibitem[Gary et al.(2006)]{gar06} Gary, S. P., Yin, L., \& Winske, D. 2006, J. Geophys. Res.
111, A06105

\bibitem[Gary et al.(2008)]{gar08} Gary, S. P., Saito, S., \& Li, H. 2008, Geophys. Res. 
Lett. 35, L02104

\bibitem[Kasper et al.(2008)]{kas08} Kasper, J. C., Lazarus, A. J., \& Gary, S.
P. 2008, Phys. Rev. Lett., 101, 261103

\bibitem[Kolmogorov(1941)]{kol41} Kolmogorov, A. N. 1941, Dokl. Akad. Nauk SSSR 
30, 9

\bibitem[Leamon et al., 2000]{leamon00} Leamon R. J. , Matthaeus W. H., Smith C. W., Zank G. P., Mullan D. J., \& Oughton S. 2000, ApJ, 537, 1054
.
\bibitem[Mangeney et al.(2002)]{man02} Mangeney, A., Califano, F., Cavazzoni, 
C., \& Tr\'{a}vn\'{i}\v{c}ek, P. 2002, J. Comput. Phys., 179, 405

\bibitem[Markovskii \& Vasquez(2002)]{mark2011} 
Markovskii, S. A., \& Vasquez, B. J. 2011, Astrophys. J., 739, 22.


\bibitem[Marsch et al.(1982a)]{mar82_a} Marsch, E., M\"{u}hlh\"{a}user K.-H.,
Rosenbauer H., Schwenn R., \& Neubauer, F. M. 1982a, J. Geophys. Res.,
87, A1, 35

\bibitem[Marsch et al.(1982b)]{mar82_b} Marsch, E., M\"{u}hlh\"{a}user K.-H.,
Schwenn R., Rosenbauer H., Pilipp W., \& Neubauer, F. M. 1982b, J. Geophys.
Res., 87, A1, 52

\bibitem[Marsch(2006)]{mar06} Marsch, E. 2006, Living Rev. Solar Phys., 3, 1

\bibitem[Marsch(2010)]{mar10} 
Marsch, E. 2010, Space Sci Rev, doi:10.1007/s11214-010-9734-z

\bibitem[Maruca et al.(2012)]{maruca12} 
Maruca, B. A., Kasper, J. C., \& Gary, S. P. 2012, Astrophys. J. 748, 137


\bibitem[Mininni \& Pouquet(2009)]{min09} Mininni, P. D., \& Pouquet, A. 2009,
Phys. Rev. E, 80, 025401

\bibitem[Osman et al.(2011)]{osm10}
Osman, K. T., Matthaeus, W. H., Greco, A., \& Servidio, S. 2011, Astrophys. J. Lett. 727, L11

\bibitem[Osman et al.(2012)]{osm12} 
Osman, K. T., Matthaeus, W. H., Hnat, B., \& Chapman, S. C. 2012, Phys. Rev. Lett. 108, 261103

\bibitem[Parashar et al.(2010)]{par10} 
Parashar, T. N., Servidio, S., Breech, B., Shay, M. A., \& Matthaeus, W. H. 2010, Phys. Pasmas 17, 102304

\bibitem[Parashar et al.(2011)]{par11} Parashar, T. N., Servidio, S., Shay, M. A., 
Breech, B., \& Matthaeus, W. H. 2011, Phys. Pasmas 18, 092302

\bibitem[Perrone et al.(2011)]{per11} Perrone, D., Valentini F., \& Veltri P.
2011, ApJ, 741, 43

\bibitem[Peyret \& Taylor(1986)]{pey86} Peyret, R., \& Taylor, T., D. 1986, 
Computational methods for fluid flow, Springer-Verlag, New-York Heidelberg
Berlin

\bibitem[Podesta & Gary(2011)]{pod11} Podesta, J. J., \& Gary, S. P. 2011, ApJ, 742, 41

\bibitem[Sahraoui et al.(2010)]{sah10} Sahraoui, F., Goldstein, M. L., Belmont,
G., Canu, P., \& Rezeau, L. 2010, Phys. Rev. Lett., 105, 131101

\bibitem[Saito et al.(2008)]{sai08} Saito, S.,  Gary, S. P., Li, H., \& Narita, Y. 2008, 
Phys. Plasmas 15, 102305 


\bibitem[Servidio et al.(2009)]{ser09} Servidio, S., Matthaeus, W. H., Shay,
M. A., Cassak, P. A., \& Dmitruk, P. 2009, Phys. Rev. Lett., 102, 115003

\bibitem[Servidio et al.(2012)]{ser12} Servidio, S., Valentini, F., Califano,
F., \& Veltri, P. 2012, Phys. Rev. Lett., 108, 045001

\bibitem[Valentini et al.(2005)]{val05} Valentini, F., Veltri, P., \& Mangeney, A. 2005, 
J. Comput. Phys., 210, 730

\bibitem[Valentini et al.(2007)]{val07} Valentini, F., Tr\'{a}vn\'{\i}\v{c}ek, 
P., Califano, F., Hellinger, P., \& Mangeney, A. 2007, J. Comput. Phys., 225,
753

\bibitem[Valentini et al.(2008)]{val08} Valentini, F., Veltri, P., Califano, F.,
\& Mangeney, A. 2008, Phys. Rev. Lett., 101, 025006

\bibitem[Valentini \& Veltri(2009)]{val09} Valentini, F., \& Veltri, P. 2009,
Phys. Rev. Lett., 102, 225001

\bibitem[Valentini et al.(2010)]{val10} Valentini, F., Califano, F., \&
Veltri, P. 2010b, Phys. Rev. Lett., 104, 205002

\bibitem[Valentini et al.(2011)]{val11} Valentini, F., Perrone, D., \&
Veltri, P. 2011, ApJ, 739, 54



\end{thebibliography}
\end{document}